\documentclass[12pt,psfig]{article}
\usepackage{graphicx}
\begin{document}

\begin{center} 
{\bf  The mean free path of protons and neutrons \\          
in isospin-asymmetric nuclear matter   }              
\end{center}
\vspace{0.1cm} 
\begin{center} 
 F. Sammarruca \\ 
\vspace{0.2cm} 
 Physics Department, University of Idaho, Moscow, ID 83844, U.S.A   
\end{center} 
\begin{abstract}
We calculate the mean free path of neutrons and protons in symmetric and 
asymmetric nuclear matter, based on  microscopic in-medium 
nucleon-nucleon cross sections. The latter are obtained from                  
calculations of the $G$-matrix including relativistic ``Dirac" effects. 
The dependence of the mean free path on energy and isospin asymmetry              
is discussed. We conclude by suggesting possible ways our
microscopic predictions might be helpful in conjunction with studies of
rare isotopes. 
\\ \\ 
PACS number(s): 21.65.+f,21.30.Fe
\end{abstract}

\section{Introduction} 

Previously, we reported 
microscopic predictions of effective nucleon-nucleon (NN) cross sections in isospin symmetric and 
asymmetric nuclear matter \cite{sk05}. In asymmetric matter, cross sections become
isospin dependent beyond the usual and well-known differences between the basic $np$, $pp$, and $nn$ 
interactions. They depend upon the total density 
and the relative proton and neutron concentrations, which implies that the 
$pp$ and the $nn$ cases will in general be different from each other.                  

In-medium cross sections are a way to explore  
the effective NN interaction in a dense and isospin-asymmetric hadronic 
environment.                     
This environment can be produced in the laboratory via energetic
heavy ion collisions (HIC).                                                       
Transport equations, such as the Boltzmann-Uehling-Uhlenbeck (BUU) equation, 
describe the evolution of a non-equilibrium gas of 
strongly interacting hadrons drifting                
in the presence of the mean field while undergoing    
two-body collisions. Thus HIC simulations require the                                      
knowledge of in-medium two-body cross sections as well as the mean field.
In a microscopic approach, both are calculated 
self-consistently starting from the bare two-nucleon force.

Besides being a crucial part of the input for transport models, in-medium effective
cross sections are important in their own right as they allow to establish an immediate connection
with the nucleon mean free path, $\lambda$, one of the most fundamental properties characterizing the 
propagation of nucleons through matter. 
The mean free path enters the calculation of the nuclear 
transparency function. The latter is obviously related to the total reaction 
cross section of a nucleus, which can be used to extract
 nuclear r.m.s. radii within Glauber-type models \cite{Glauber}. Therefore,  microscopic in-medium 
 isospin-dependent NN cross sections can ultimately help obtain 
information about the properties of exotic, neutron-rich nuclei. 
These studies are particularly timely due to 
the advent of radioactive beams, which allow to explore                            
the unknown regions of proton/neutron rich 
unstable nuclei.                                                                             

Applying our microscopic cross sections in calculations of the nucleon mean free
path in symmetric and asymmetric nuclear matter is the focal point of this note. 
Recently, predictions of the mean free path have been obtained
from the nucleon optical potential calculated in the relativistic impulse approximation, 
together with empirical NN scattering amplitudes and the relativistic mean field model  
\cite{Jiang07}.
Those were then used to extract in-medium cross sections. 
Our calculations are microscopic and proceed exactly in the opposite way,  namely we obtain    
$\lambda$ from the microscopically predicted cross sections. It will be interesting to see
if some consistency can be found between the two sets of results. 

In the next section, we recall the main aspects 
of the previously calculated cross sections, which, together with neutron and proton densities, 
completely determine the mean free path. We then present 
and discuss our results in Section III. Our conclusions and outlook are summarized in Section IV.

\section{Effective cross sections and mean free path}                
Our cross sections are calculated from a $G$-matrix which includes all ``conventional'' medium effects as well as 
those associated with medium modifications of the nucleon Dirac wavefunction 
(DBHF effects). We choose 
the Bonn-B potential \cite{Mac89} as our model for the free-space 
two-nucleon force.                           
 The nuclear matter calculation of Ref.~\cite{AS02} provides, 
self-consistently with the nuclear equation of state, 
the single-proton/neutron potentials as well as their parametrizations in 
terms of effective masses. Those effective masses, together with 
the appropriate Pauli operator (depending on the type of nucleon involved), 
 are then used in a separate calculation of the in-medium       
 reaction matrix (or $G$-matrix) under the desired kinematical conditions.
Coulomb effects are not included in the $pp$ cross sections, which therefore differ from the 
$nn$ ones entirely due to the proton and the neutron having different Fermi momenta. 
In Ref.~\cite{sk05} we found that 
the degree of sensitivity to the asymmetry in neutron and proton concentrations depends strongly on the 
region of the energy-density-asymmetry phase space under consideration, 
and can separate $pp$ and $nn$ scatterings under appropriate conditions of density
and kinematics. 

We recall that the neutron and proton Fermi momenta, 
$k_F^n$ and
$k_F^p$, change with increasing neutron 
fraction according to the relations
\begin{equation}
k_F^n = k_F(1 + \alpha)^{1/3}
\end{equation}
\begin{equation}
k_F^p = k_F(1 - \alpha)^{1/3} , 
\end{equation}
where $k_F$ is the average Fermi momentum, 
 and $\alpha=(\rho_n - \rho _p)/(\rho_n + \rho_p)$.

In Ref.~\cite{sk05} we calculated
the total cross section as                   
\begin{equation}
\sigma(q_0,P_{tot},\rho)= \int \frac{d\sigma}{d\Omega}                               
Q(q_0,P_{tot},\theta, \rho)d\Omega , 
\end{equation}
where
$\frac{d\sigma}{d\Omega} $                                    
is given by the usual sum of amplitudes squared and phase space factors and      
$Q$ is the Pauli operator which prevents scattering into occupied states.

The momentum $q_0$ is                                   
the two-body c.m. frame momentum, and                 
$P_{tot}$ is the total momentum 
of the two-nucleon system, which, in the present calculation, is taken to be             
equal to zero. This amounts to assuming  
that the c.m. frame of the nucleons
and the nuclear matter rest frame coincide, a                                                  
choice which is different from the one employed in 
 Ref.~\cite{sk05} but
consistent with how we calculated the cross sections in Ref.~\cite{Lombardo}.                

Notice that the Pauli operator in Eq.~(3) is present in addition to Pauli blocking                         
of the (virtual) intermediate states. The latter acts only on the intermediate states by 
cutting out part of the momentum
spectrum during the integration of the scattering equation, whereas Eq.~(3) prevents scattering      
into occupied final states.                                                   
In the present case, because we are taking 
$P_{tot}$ to be equal to zero, and we are                        
considering elastic scattering, the presence of $Q$ in Eq.~(3)                    
is equivalent to simply setting the cross section
to zero whenever the momentum $q_0$                                        
is below the Fermi level, since 
in such case the scattering is forbidden. In other words,                                        
our momenta and energies are defined relative to the bottom of the Fermi sea, and we have in mind a  
scenario where 
a nucleon is bound in a nucleus (or, more ideally, nuclear matter) through the mean field. If such 
nucleon is struck, (for instance, in a $(e,e^{'})$ reaction), it may subsequently              
 scatter from another nucleon.

\begin{figure}
\begin{center}
\vspace*{-4.0cm}
\hspace*{-2.0cm}
\scalebox{0.4}{\includegraphics{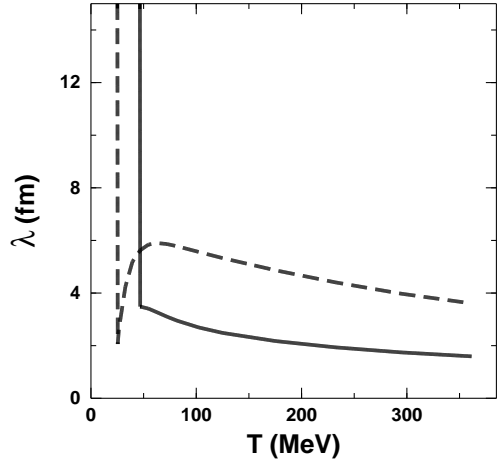}}
\vspace*{-3.0cm}
\caption{Nucleon mean free path in symmetric nuclear matter as a function 
of the nucleon kinetic energy at an average Fermi momentum of 1.1$fm^{-1}$ (dash)
and 1.4$fm^{-1}$ (solid). 
} 
\label{one}
\end{center}
\end{figure}

\begin{figure}
\begin{center}
\vspace*{-4.0cm}
\hspace*{-2.0cm}
\scalebox{0.5}{\includegraphics{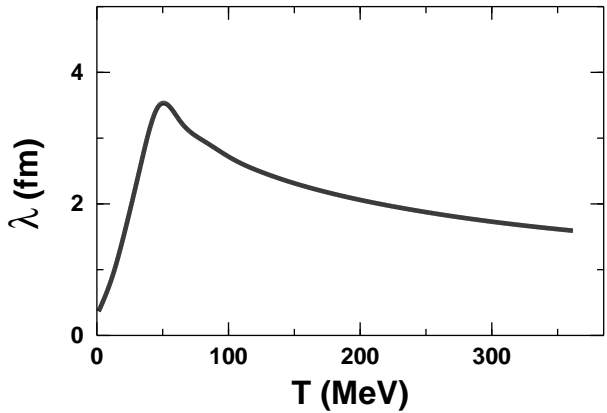}}
\vspace*{-4.0cm}
\caption{ 
Behaviour of the mean free path when the cross section is not 
suppressed by Pauli blocking of the final scattering state. The Fermi momentum 
is equal to 1.4$fm^{-1}$.                    
} 
\label{two}
\end{center}
\end{figure}

It was shown by Negele and Yazaki 
\cite{NY81} that                   
the nucleon mean free path is related to the imaginary part of the dispersion 
relation through 
\begin{equation}
\lambda = \frac{1}{2k^I}, 
\end{equation}
where $k^I$ is the imaginary part of the complex momentum.    
The mean free path can also be defined, for a proton, as
\begin{equation}
\lambda_p = \frac{1}{\rho_p \sigma_{pp} + \rho_n \sigma_{pn}}, 
\end{equation}
with $\rho_n$ and $\rho_p$ the neutron and proton densities in 
asymmetric matter, equal to 
$\frac{(k_F^n)^3}{3\pi^2}$ and 
$\frac{(k_F^p)^3}{3\pi^2}$,            
respectively. (An analogous definition holds for the neutron.) 
The above expression represents the length of the unit volume in the phase space 
defined by the effective scattering area and the number 
of particles/volume \cite{PP}. 
Notice that we set the appropriate cross section to zero if the nucleon momentum is less than or equal 
to the Fermi momentum (of that particular nucleon type), since final states 
(as well as intermediate ones) are Pauli-blocked when calculating  the mean free path for the reasons 
mentioned above. 
Equivalent considerations in 
Ref.~\cite{sk05} meant that the angular integration in Eq.~(3) was restricted through a condition
involving $P_{tot}$ as well.

\section{Results and discussion}                                     
First, we show the mean free path of nucleons in symmetric matter as a function 
of the nucleon kinetic energy (calculated as $T=\sqrt{q_0^2+m^2}-m$),                                    
see Fig.~1. The chosen densities correspond to Fermi 
momenta of 1.1$fm^{-1}$ and 1.4$fm^{-1}$ for the dashed and the solid curve, respectively. 
The density dependence is quite large. 
Again, in the present approximation, the 
cross section goes sharply to zero, and thus the mean free path goes to infinity, for 
$q_0\le k_F$. Thus                                                                      
the lowest energy for which $\lambda$ is finite corresponds to the lowest momentum
allowed by Pauli blocking of the final state. Table~1, together with Eq.~(5), 
should facilitate the interpretation 
of the mean free path behaviour  
observed in Fig.~1. 
Reconnecting with 
the previous discussion which followed Eq.~(3), we also show, see Fig.~2, 
the mean free path calculated without considerations of Pauli blocking of the final states. 
In this case, $\lambda$ becomes very small at low energy, due to the large values of the 
cross section in that region. 
                                     
Back to Fig.~1, and focusing on the higher density first (solid line), 
we see the sharp drop from infinity at low energy, after which 
the mean free path slowly decreases with energy, due to the fact that
the in-medium cross sections actually start to go up with energy at high
densities, see Table~1.                                                                        
This feature, which may apppear counterintuitive (being               
opposite to what is seen in free space), has been reported    
in other works as well \cite{Jiang07,Fuchs}. A similar behavior also sets in at the lower density (dashed line),           
but in that case the mean free path, after the sharp drop from infinity,                                      
rises with energy at first (corresponding to a reduction of the in-medium 
cross section).                                                                                      
Notice that the tendency to rise with energy 
in dense 
matter appears more pronounced for scattering of identical nucleons, a behaviour which was traced to in-medium
enhancement of some isospin-1 partial waves \cite{sk05}.

\begin{table}                
\centering \caption                                                    
{$pp$ and $np$ total effective cross sections in symmetric matter calculated at two 
densities as a function of the kinetic energy. 
} 
\vspace{5mm}

\begin{tabular}{|c|c|c|c|}
\hline

$k_{F}(fm^{-1})$ & $T(q_0)(MeV)$ & $\sigma _{pp}(mb)$ &
                   $\sigma _{np}(mb)$                                
                    \\                              
\hline
 1.1 &  5.31      &  .0000     &      .0000 \\ 
    &  8.28     &   .0000      &     .0000  \\
    &   11.91    &    .0000    &       .0000       \\
    &   16.17    &    .0000      &     .0000               \\
    &      21.06   &      .0000  &         .0000   \\ 
    &      26.58    &  23.46     &      60.57      \\
    &    32.71     & 18.00       &    34.39       \\
    &    39.44    &  16.67      &     26.64      \\
    &    46.76    &  16.41      &     23.14     \\ 
    &    54.66    &  16.63       &    21.44      \\
    &    63.11    &  17.08       &    20.63      \\ 
    &    72.12    &  17.67       &    20.28    \\ 
    &    81.65    &  18.31       &    20.17    \\ 
    &   102.27    &  19.66       &    20.32     \\ 
    &  124.83     & 21.03        &   20.71       \\ 
    &     175.34   &   23.79     &      21.94      \\ 
    &      232.22   &   26.66    &       23.83     \\ 
    &     294.60    &  29.66     &      26.30     \\ 
    &     361.68    &  32.63     &      29.08     \\ 
\hline
1.4&      5.31      &  .0000  &    .0000      \\ 
   &  8.28     &   .0000  &    .0000        \\ 
   &     11.91   &      .0000  &    .0000    \\ 
   &     16.17   &     .0000   &   .0000     \\ 
   &     21.06   &     .0000   &   .0000   \\ 
   &     26.58   &     .0000   &   .0000   \\ 
   &     32.71   &     .0000   &   .0000    \\ 
   &     39.44   &     .0000   &   .0000    \\ 
   &     46.76   &   13.70     & 17.26      \\ 
   &     54.66   &   15.04     & 16.63      \\ 
   &     63.11   &   16.31     & 16.77      \\ 
   &     72.12   &   17.54     & 17.24     \\ 
   &     81.65   &   18.71     & 17.85      \\ 
   &    102.27   &   20.86     & 19.22      \\ 
   &    124.83   &   22.80     & 20.61      \\ 
   &    175.34   &   26.26     & 23.37      \\ 
   &    232.22   &   29.47     & 26.28      \\ 
   &    294.60   &   32.55     & 29.32      \\ 
   &    361.68   &   35.40     & 32.34                 
\\                  
\end{tabular}
\end{table}

\begin{figure}
\begin{center}
\vspace*{-4.0cm}
\hspace*{-2.0cm}
\scalebox{0.4}{\includegraphics{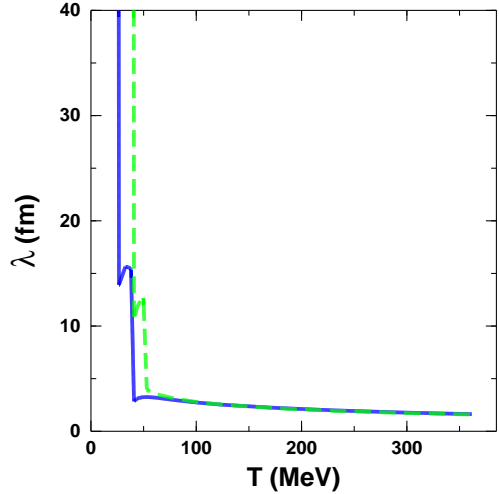}}
\vspace*{-3.0cm} 
\caption{                                   
Mean free path for neutrons (dashed) and protons (solid) for an average Fermi momentum of     
1.4$fm^{-1}$
 and isospin asymmetry, $\alpha$, equal to 0.5. 
} 
\label{three}
\end{center}
\end{figure}

\begin{figure}
\begin{center}
\vspace*{-4.0cm}
\hspace*{-2.0cm}
\scalebox{0.4}{\includegraphics{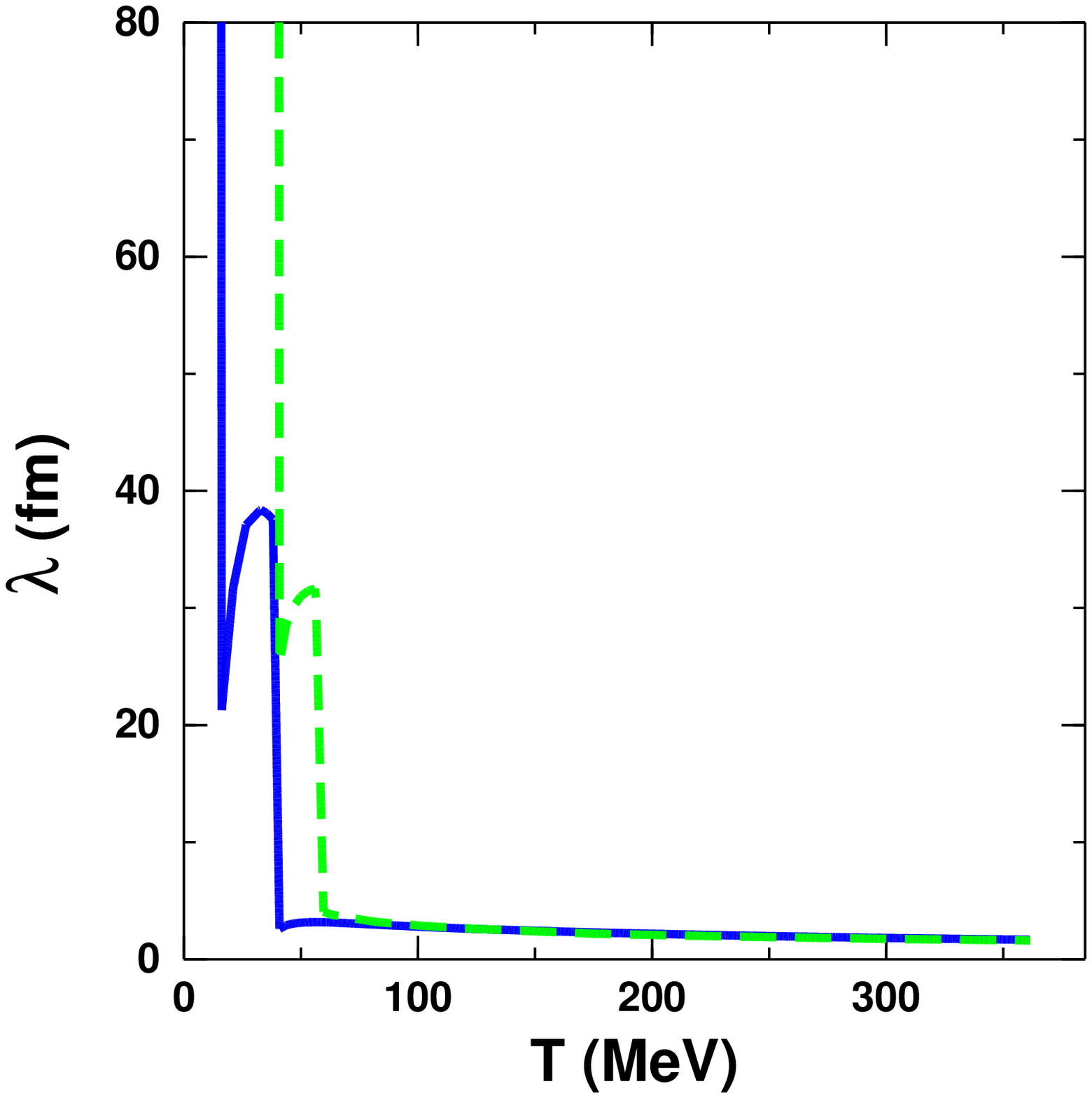}}
\vspace*{-3.0cm} 
\caption{                                   
Same as in Fig.~3, but with                                             
$\alpha$ equal to 0.8. 
} 
\label{four}
\end{center}
\end{figure}
        
We now move to mean free path considerations 
in asymmetric matter. For scattering of like nucleons, the cross section is set equal to 
zero when 
$q_0\le k_F^i$ ($i=n,p$), whereas for $np$ scattering it is set to zero for                
$q_0\le k_F$, the {\it average} Fermi momentum.                                            
The corresponding behavior of the mean free path is shown in Fig.~3 for $\alpha$=0.5 
and in Fig.~4 for a greater degree of asymmetry, $\alpha$=0.8. The large differences between the mean free path for protons  
and neutrons 
at the lowest energies are to be expected from what we stated above, namely 
the suppression of $pp$ and $nn$ cross sections is controlled by the (unequal)
proton and neutron Fermi momenta. Proceeding from the lowest to the highest energies,
the proton mean free path is infinity when both $pp$ and $np$ cross sections are Pauli blocked,
followed by the small rise around 25 MeV, and then again the sharp drop when the $np$ cross 
section starts to contribute. 
Similar considerations explain the dashed curve, with the difference that the neutron Fermi momentum 
is the highest in this case. 
These effects are of course especially pronounced when $k_F^n$ is much larger than
$k_F^p$, see Fig.~4.                             

In summary, strong variations with energy of the proton and neutron mean free path can be seen, as       
well as large differences between the two, in a rather narrow region around the Fermi 
``thresholds" for $pp$ and $nn$ scatterings. 
As energy increases, however, the mean free path becomes essentially insensitive to isospin 
asymmetry. This is in agreement with the conclusions of Ref.~\cite{Jiang07}. 

\begin{table}                
\centering \caption                                                    
{$pp$, $nn$, and $np$ total effective cross sections in asymmetric matter under the same 
conditions as chosen in Fig.~3. 
} 
\vspace{5mm}

\begin{tabular}{|c|c|c|c|}
\hline
$T(q_0)(MeV)$ & $\sigma_{pp}(mb)$ & $\sigma_{nn}(mb)$ & $\sigma _{np}(mb)$  
\\                  \\                              
\hline
      5.31   &     .0000  &    .0000 &     .000  \\ 
      8.28   &     .0000  &    .0000 &     .0000 \\ 
     11.91   &     .0000  &    .0000 &     .0000   \\ 
     16.17   &     .0000  &    .0000 &     .0000   \\ 
     21.06   &     .0000  &    .0000 &     .0000  \\ 
     26.58   &   15.51    &    .0000 &     .0000  \\ 
     32.71   &   13.80    &    .0000 &     .0000   \\ 
     35.33   &   13.82    &    .0000 &     .0000   \\ 
     38.05   &   14.00    &    .0000 &     .0000   \\ 
     40.86   &   14.27    &    .0000 &   20.50     \\ 
     43.77   &   14.62    &    .0000 &   18.32     \\ 
     46.76   &   15.01    &    .0000 &   17.48     \\ 
     49.85   &   15.45    &    .0000 &   17.04     \\ 
     53.03   &   15.90    &  12.52   &   16.83     \\ 
     56.30   &   16.37    &  13.54   &   16.76      \\ 
     59.66   &   16.85    &  14.29   &   16.78     \\ 
     63.11   &   17.33    &  14.92   &   16.87     \\ 
     81.65   &   19.70    &  17.51   &   17.90     \\ 
    102.27   &   21.85    &  19.68   &   19.22     \\ 
    124.83   &   23.79    &  21.61   &   20.57      \\ 
    175.34   &   27.20    &  25.09   &   23.29     \\ 
    232.22   &   30.34    &  28.37   &   26.16     \\ 
    294.60   &   33.32    &  31.53   &   29.19     \\ 
    361.68   &   36.06    &  34.50   &   32.21  
\\ 
\hline

\end{tabular}
\end{table}

Finally, we show in Table~2 some of the in-medium cross sections used for the present
calculations of the mean free path in asymmetric matter. The $pp$ and $nn$ cross sections are quite similar to 
each other except in the low energy region where one may be sizable while the other is
still suppressed.

\section{Conclusions and future prospects}
We presented predictions of the mean free path for protons and neutrons in isospin
symmetric or asymmetric matter based on microscpic predictions of in-medium
cross sections. The mean free path 
in exotic matter is a fundamentally important quantity which finds applications
in diverse areas including radiobiology. 

As it appears reasonable,                  
very low-energy protons and neutrons can have dramatically different propagation properties in 
strongly asymmetric matter.                                                                               
Our conclusion is that 
an experimental signature of sensitivity of in-medium scattering to isospin asymmetry
may be sought by probing highly asymmetric matter with energies close to the proton and neutron    
Fermi surfaces. 
Otherwise, isospin asymmetry has only a very minor impact on the mean free path. 

We recall that 
our baseline calculation of the cross sections is a microscopic one. The assumptions we made in this 
paper concerning kinematics and sharpness of the Pauli operator simply have the purpose to make the   
discussion 
more transparent and can be improved or removed depending on the specific needs of potential users and 
the experimental conditions one may wish to simulate. 

Through additional steps, which would involve the calculation of the 
nuclear trasparency function (defined as the probability that at some impact parameter
the projectile will pass through the target without interacting), 
the mean free path is closely related 
to the nuclear reaction cross section.                               
Thus, analyses of reaction cross section data 
can ultimately shed light on the target density (a much needed information for
nuclei with large neutron skin and, thus, hard-to-probe density distributions).
On the other hand, 
a crucial input for the equations written above are the two-body cross sections, for which
parametrizations of free-space NN cross sections are often adopted. This 
may not be reliable, and we suggest that keeping in touch with 
microscopic predictions can be of help when trying to constrain observables which depend on 
several (essentially unknown) degrees of freedom.

\begin{center}
{\bf ACKNOWLEDGMENTS}
\end{center}
Financial support from the U.S. Department of Energy under grant number DE-FG02-03ER41270 
is acknowledged.

\end{document}